\documentclass[11pt,a4paper]{article}
\usepackage{amsmath,amssymb}
\usepackage{framed}
\usepackage{fullpage}
\usepackage{hyperref}
\usepackage{graphicx}
\usepackage{amsfonts,amsthm,url,tabularx, array}
\usepackage{enumerate,verbatim,float} 
\usepackage{framed}
\usepackage{bold-extra}

\DeclareMathOperator{\layer}{\mathsf{layer}}

\DeclareMathOperator{\FO}{FO}
\DeclareMathOperator{\MSO}{MSO}
\DeclareMathOperator{\arity}{arity}
\DeclareMathOperator{\tw}{tw}
\DeclareMathOperator{\dom}{dom}
\DeclareMathOperator{\inc}{inc}

\newcommand{\emb}{$p$-\textsc{Emb}}
\newcommand{\embcount}{$p$-\#\textsc{Emb}}

\newtheorem{thm}{Theorem}

\newtheorem{lma}[thm]{Lemma}

\newtheorem{cor}[thm]{Corollary}

\newtheorem*{adef}{Definition}

\title{Two dichotomies for model-checking in multi-layer structures}
\author{Jessica Enright \and Kitty Meeks \and Jessica Ryan \\
\small{School of Computing Science, University of Glasgow, Glasgow, UK\thanks{Email addresses: \texttt{jessica.enright@glasgow.ac.uk}, \texttt{kitty.meeks@glasgow.ac.uk}, \texttt{j.ryan.2@research.gla.ac.uk}}}}

\begin{document}

\date{June 2020}
\maketitle

\begin{abstract} 
Multi-layer graphs can capture qualitatively different types of connection between entities, and networks of this kind are prevalent in biological and social systems: for example, a social contact network typically involves both virtual and face-to-face interactions between individuals.  Since each layer is likely to exhibit stronger and/or more easily identifiable structural properties than the overall system, it is natural to ask whether we can exploit the structural properties of individual layers to solve NP-hard problems efficiently on the overall network.  In this paper we provide a complete characterisation of the structural properties required in each layer to guarantee the existence of an FPT algorithm to solve problems definable in either first-order or monadic second-order logic on the overall system, subject to the assumption that the structural properties are preserved under deletion of vertices and/or edges.\\
\vspace{8pt}
\end{abstract}

\section{Introduction}

A multi-layer (or multiplex) network includes edges that may be qualitatively different, and describe different types of interaction: for example, layers might correspond to different varieties of social interaction, or physical as compared to electronic contact \cite{kivelaReview}.  The capacity of multi-layer networks to represent  physical and social systems has made their study one of the leading areas of research in network science \cite{kivelaReview,pilosofPorter}, although as yet there are only a few algorithmic results concerning ``layered'' graph problems \cite{brederek19,kratsch18}.  Understanding the multi-layer nature of many real-world inputs is important in the design of algorithms to solve NP-hard problems, even if our goal is to answer questions about the ``flattened'' graph formed by combining the edges from all layers: each individual layer is likely to exhibit stronger and more easily understandable structure than the combined graph, and we may hope to leverage such structure to develop efficient algorithms.  For example, when considering contact between individuals in a population of livestock with a view to modelling or controlling the spread of disease, the network of contacts has  at least two distinct types of connection: those due to geographic proximity (farms that share a common border) and those that arise from trade.  The geographic layer will necessarily be planar with reasonably low degree, whereas previous work suggests that the trade layer is likely to have low treewidth \cite{edgeDeletion}, and only those vertices representing markets or dealers are likely to have very large degree.  

We are therefore interested in understanding the conditions under which algorithmically useful structure in each layer can be exploited to design efficient algorithms to solve problems on the entire system. In this paper we provide a complete characterisation of the structural conditions that must be imposed on each layer to allow the design of FPT algorithms for both decision and counting problems on the overall structure that are definable in first-order or monadic second-order logic, provided that the structural properties under consideration are monotone, i.e. are preserved under the deletion of vertices and/or edges.  A huge number of practically important graph problems are definable in one of these two fragments of logic: for example, the subgraph isomorphism problem (widely used in the analysis of social and biological networks \cite{schank05,tsourakakis08,milo02,guillemot13,  connected}) is definable in first-order logic, while the extension to monadic second-order logic allows us to consider graph modification problems such as those that have already been studied in individual layers of the livestock contact network described above \cite{edgeDeletion}.  The complexity of model-checking and counting for both first-order and monadic second-order logic on single-layer graphs is already well understood, but this is the first work to address the complexity of these problems explicitly in the multi-layer setting.  While the initial motivation comes from the study of graph problems, it is convenient to work in the more general framework of relational structures, which provides a unified way of representing graphs equipped with some additional information such as a colouring of the vertices or edges.  

While all of our results are obtained using fairly standard techniques, this is the first time such techniques have been applied in the context of multi-layer networks, and it is somewhat surprising that we are able to obtain a complete characterisation with elementary techniques.  The strength of our hardness results for the two settings provides motivation for the development of more complex models for the structure of multi-layer graphs which restrict not only the structure of individual layers but also the interactions between the layers; we anticipate that this will be a rich direction for future research.

The rest of the paper is organised as follows.  We begin in Section \ref{sec:notation} with the necessary definitions and background.  In Section \ref{sec:layers} we give the formal definition of our model for layered graphs and structures, and prove a number of facts about subgraphs which can be found in ``layered'' graph classes when the layers have specific properties; these results are used later to prove our hardness results.  Our two dichotomy results are stated and proved in Section \ref{sec:dichotomies}.

\section{Preliminaries}
\label{sec:notation}

\paragraph*{Sets, graphs and graph parameters} 

Given any natural number $r$, we write $[r]$ as a shorthand for $\{1,\ldots,r\}$.  If $\pi$ is a function with domain $X$, and $Y \subseteq X$, we write $\pi(Y)$ for the image of $Y$ under $\pi$.

A graph $G$ is a pair $(V,E)$ where the \emph{vertex set} $V = V(G)$ is any finite set and the \emph{edge set} $E = E(G)$ is a subset of the set of unordered pairs of elements of $V$; we write $uv$ for the edge consisting of $u$ and $v$, and say that $u$ and $v$ are \emph{adjacent}.  Given any graph $G = (V,E)$, and a vertex $v \in V$, we write $d_G(v)$ for the degree of $v$ in $G$.  
Given a subset $U \subset V$, we write $G[U]$ for the subgraph of $G$ induced by $U$, and $G \setminus U$ for the subgraph obtained from $G$ by deleting all elements of $U$.  If $F \subseteq E(G)$, we denote by $G \setminus F$ the graph obtained from $G$ by deleting all edges in $F$.  We write $E_G(U,*)$ for the set of edges in $G$ with at least one endpoint in $U$ (omitting the subscript if it the graph $G$ is clear from the context).  

For $u,v \in V$, the \emph{distance} between $u$ and $v$ in $G$ is the number of edges on a shortest path between $u$ and $v$ in $G$.  A star is a graph isomorphic to the complete bipartite graph $K_{1,p}$ for some $p \in \mathbb{N}$.  A \emph{star forest} is an acyclic graph in which every connected component is a star.  A $k$-clique in a graph $G$ is a subgraph of $G$ consisting of $k$ pairwise adjacent vertices.

\begin{figure}
\centering
\includegraphics[width = 0.6 \linewidth]{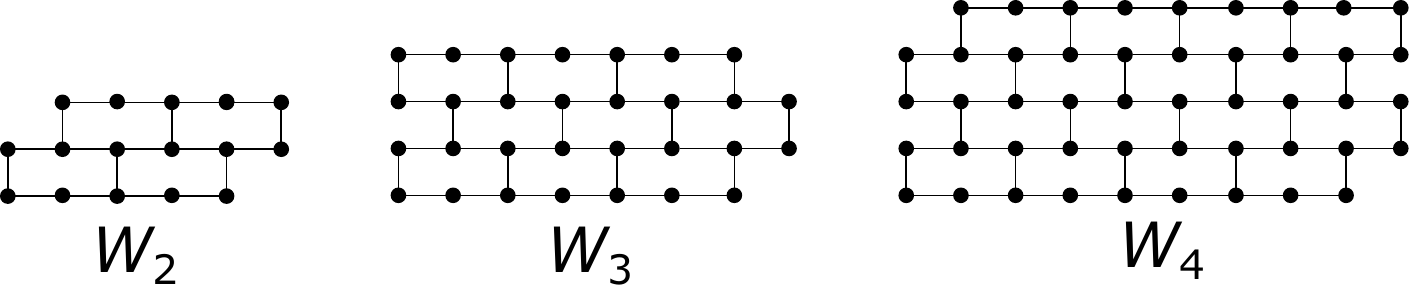}
\caption{Walls of height two, three and four.}
\label{fig:wall-egs}
\end{figure}

One specific family of graphs will be particularly useful in proving our results.  The wall of height $h$ is the graph $W_r = (V_r,E_r)$, where
\begin{equation*}
V_r = \begin{cases}
			\{(i, j) | i \in [h + 1], j \in [2h + 2]\} \setminus \{(1,2h+2),(h+1,2h+2)\}		& \text{if $h$ is odd,}\\
			\{(i, j) | i \in [h + 1], j \in [2h + 2]\} \setminus \{(1,2h+2),(h+1,1)\}			& \text{if $h$ is even,}
	  \end{cases}
\end{equation*}
and vertices $(i,j)$ and $(i',j')$ are adjacent if and only if either
\begin{itemize}
\item $i=i'$ and $|j - j'| = 1$ (which gives a \emph{horizontal edge}), or
\item $j = j'$ and $i' = i + (-1)^{i+j}$ (which gives a \emph{vertical edge}).
\end{itemize}
Walls of height two, three and four are illustrated in Figure \ref{fig:wall-egs}.  It is well-known that the wall $W_h$ has treewidth exactly $h$ (folklore theorem). 

A graph class is said to be \emph{monotone} if it is closed under the deletion of both vertices and edges.  A set $X \subset V$ is a \emph{vertex cover} for $G$ if $G \setminus X$ is an independent set (i.e. contains no edges), or equivalently if $E = E_G(X,*)$; the \emph{vertex cover number} of $G$ is the cardinality of the smallest vertex cover for $G$.  We say that a class $\mathcal{C}$ of graphs has bounded vertex cover number if there exists a constant $c$ such that every graph in $\mathcal{C}$ has vertex cover number at most $c$.

Given any graph class $\mathcal{C}$, we say that another graph class $\mathcal{C}'$ is \emph{at constant vertex cover distance from $\mathcal{C}$} if there exists a constant $c$ such that, for every graph $G \in \mathcal{C}'$, there is a set $X \subseteq V(G)$ such that $|X| \leq c$ and $G \setminus E_G(X,*) \in \mathcal{C}$; we say that such a set $X$ \emph{witnesses} the vertex cover distance.  Thus the class of graphs with bounded vertex cover number is the class of graphs at constant vertex cover distance from the class of all edgeless graphs.

We assume throughout the paper that all graph classes are isomorphism-closed, that is if $G \in \mathcal{C}$ and $H$ is isomorphic to $G$, then also $H \in \mathcal{C}$.

\paragraph*{Relational structures}  

We begin by recalling some standard terminology for describing relational structures; for further background we refer the reader to \cite[Chapter 4]{flumgrohe}.  A \emph{vocabulary} is a finite set $\tau$ of relation symbols, where each relation symbol $R$ is associated with a natural number, $\arity(R)$, known as its \emph{arity}.  For any vocabulary $\tau$, a \emph{$\tau$-structure} is a pair $\mathfrak{A} = (A, \{R^{\mathfrak{A}}: R \in \tau\})$ such that, for each $R \in \tau$, if $R$ has arity $r$ then $R^{\mathfrak{A}} \subseteq A^r$.  We call $A$ the \emph{universe} of $\mathfrak{A}$, and $R^{\mathfrak{A}}$ the \emph{interpretation} of $R$ in $\mathfrak{A}$.  We synonymously write $\mathbf{a} \in R^{\mathfrak{A}}$ or $R^{\mathfrak{A}}\mathbf{a}$ to indicate that the tuple $\mathbf{a}$ belongs to $R^{\mathfrak{A}}$.  In general a $\tau$-structure may be infinite, but here we consider only finite structures.  

An (undirected) graph $G = (V,E)$ is a special case of a relational structure, and there are two common ways to encode a graph as a relational structure: the first has universe $V$ and a single symmetric, binary \emph{adjacency} relation $E^G$ such that $(u,v) \in E^G$ if and only if $uv$ is an edge of $G$; the second has universe $V \cup E$ and a single binary \emph{incidence} relation $I^G$ such that $(v,e) \in I^G$ if and only if $v$ is an endpoint of $e$.

The \emph{Gaifman graph} of a $\tau$-structure $\mathfrak{A}$ with universe $A$ is a graph $G = (A,E)$ where $ab \in E$ if and only if, for some positive integer $r$, there exists a relation $R \in \tau$ of arity $r$ and $(a_1,\ldots,a_r) \in R^{\mathfrak{A}}$ such that $a,b \in \{a_1,\ldots,a_r\}$; this means that two elements of the universe are connected by an edge if and only if they belong to the same tuple of at least one relation in $\mathfrak{A}$.  Note that, if the structure $\mathfrak{A}$ encodes a graph $G$ using an adjacency relation, then the Gaifman graph of $\mathfrak{A}$ is isomorphic to $G$ (but this is not true when the incidence relation is used).  For convenience, we will sometimes refer to a structure $\mathfrak{A}$ belonging to some graph class $\mathcal{C}$ (e.g. the class of graphs of bounded vertex cover number); by this we mean that the Gaifman graph of $\mathfrak{A}$ belongs to $\mathcal{C}$.

\paragraph*{First-order and monadic second-order logic} 

For any vocabulary $\tau$, the set $\FO[\tau]$ of \emph{first-order formulas} is built up from a countably infinite set of variables $x_1,x_2,\ldots$, the relation symbols $R \in \tau$, the connectives $\wedge, \vee, \neg$ and the quantifiers $\forall x, \exists x$ ranging over elements of the universe of the structure.  Given a first-order formula $\phi$, a variable $x$ appearing in $\phi$ is said to be a \emph{free variable} if $x$ is not in the scope of a quantifier $\exists x$ or $\forall x$.  We write $\phi(x_1,\ldots,x_t)$ for a formula $\phi$ with free variables $x_1,\ldots,x_t$.

Given a $\tau$-structure $\mathfrak{A}$ with universe $A$, a formula $\phi(x_1,\ldots,x_t) \in \FO[\tau]$, and elements $a_1,\ldots,a_t \in A$, we write $\mathfrak{A} \models \phi(a_1,\ldots,a_t)$ to say that $\mathfrak{A}$ satisfies $\phi$ if the variables $x_1,\ldots,x_t$ are interpreted as $a_1,\ldots,a_t$.  For a structure $\mathfrak{A}$ and a formula $\phi$, we set $\phi(\mathfrak{A}) = \{(a_1,\ldots,a_t) \in A^t \colon \mathfrak{A} \models \phi(a_1,\ldots,a_t)\}$.  This definition can also be extended to formulas with no free variables: in this case $\phi(\mathfrak{A})$ is a set containing only the empty tuple if $\mathfrak{A} \models \phi$, and the empty set otherwise.

\emph{Monadic second-order logic} additionally allows quantification over subsets of the universe; formally, this is achieved with the use of unary relation variables (each of which encodes inclusion in a subset).  We write $\MSO[\tau]$ for the set of monadic second-order formulas with vocabulary $\tau$.  Similarly to the case for first-order logic, if $\mathfrak{A}$ is a $\tau$-structure and $\phi(X_1,\ldots,X_s,x_1,\ldots,x_t) \in \MSO[\tau]$ a formula with free relation variables $X_1,\ldots,X_s$ and individual variables $x_1,\ldots,x_t$, we write $\mathfrak{A} \models \phi(A_1,\ldots,A_s,a_1,\ldots,a_t)$ to say that $\mathfrak{A}$ satisfies $\phi$ if the variables $X_1,\ldots,X_s,x_1,\ldots,x_t$ are interpreted as $A_1,\ldots,A_s,a_1,\ldots,a_t$ respectively.  In this setting, we have
\[\phi(\mathfrak{A}) = \{(A_1,\ldots,A_s,a_1,\ldots,a_t): A_1,\ldots,A_s \subseteq A, a_1,\ldots,a_t \in A, \mathfrak{A} \models \phi(A_1,\ldots,A_s,a_1,\ldots,a_t)\}.\]  
We extend to formulas with no free variables in exactly the same way as before.

\paragraph*{Parameterised complexity}  

We are interested in determining the circumstances under which the problems we consider admit FPT algorithms, that is, algorithms running in time $f(k) \cdot n^c$ where $n$ is the total input size, $k$ is the parameter, $f$ is any (computable) function, and $c$ is a fixed constant that does not depend on $k$. A much weaker requirement on the running time of an algorithm is that it is polynomial-time whenever $k$ is bounded by a fixed constant (so that, for example, a running time of $n^k$ is acceptable); problems admitting an algorithm of this kind are said to belong to the class XP.

In order to demonstrate that a decision problem is unlikely to admit an FPT algorithm, it suffices to demonstrate that it is complete for the complexity class W[1].  The corresponding complexity class for parameterised counting problems is \#W[1].  For further background on the theory of parameterised complexity we refer the reader to \cite{downeyfellows13,flumgrohe}.

In order to consider structures and formulas as the inputs to computational problems, we need a notion of their size.  A $\tau$-structure $\mathfrak{A}$ can be encoded as a string $\langle \mathfrak{A} \rangle$ whose length $|\langle \mathfrak{A} \rangle
|$ is within a polynomial factor of $|\tau| + |A| + \sum_{R \in \tau} \big( |R^{\mathfrak{A}}| \cdot \arity(R) \big)$.  A formula $\phi \in \FO[\tau]$ can be encoded by a string $\langle \phi \rangle$ with $|\phi| \leq |\langle \phi \rangle| = \mathcal{O}(|\phi| \cdot \log |\phi|)$.  Details of such encodings can be found in \cite[Chapter 4]{flumgrohe}.  We are interested in the setting in which the size of the formula may be much smaller than the size of the structure, so we consider the complexity of these model-checking problems when parameterised by the size of the formula.

\paragraph*{Problems considered}  

We consider both the decision and counting versions of model-checking for first-order and monadic second-order logic.  For a class $\Phi$ of formulas, the model-checking and counting problems are formally defined as follows.

\begin{framed}
\noindent
\textsc{$\Phi$-Model-Checking}\\
\textit{Input:} A $\tau$-structure $\mathfrak{A}$ and a formula $\phi \in \Phi[\tau]$\\
\textit{Parameter:} $|\langle \phi \rangle|$\\
\textit{Question:} Is $\phi(\mathfrak{A})$ non-empty?
\end{framed}

\begin{framed}
\noindent
\textsc{$\Phi$-Counting}\\
\textit{Input:} A $\tau$-structure $\mathfrak{A}$ and a formula $\phi \in \Phi[\tau]$\\
\textit{Parameter:} $|\langle \phi \rangle|$\\
\textit{Question:} What is $|\phi(\mathfrak{A})|$?
\end{framed}

We consider \textsc{FO-Model-Checking}, \textsc{FO-Counting}, \textsc{MSO-Model-Checking} and \textsc{MSO-Counting}.  The complexity of all four problems has been studied thoroughly in the case of single-layer structures, and there are a number of meta-theorems giving tractability for various classes.  In the first-order case, it is known that even the counting version is in FPT for large families of both sparse and dense graphs (note that tractability of the counting version immediately implies tractability for the decision problem); specifically, we have the following two results.

\begin{thm}[\cite{courcelle01}]\label{thm:clique-width}
Let $\mathcal{C}$ be a class of structures of bounded cliquewidth.  Then \textsc{FO-Counting} is in FPT when restricted to structures from $\mathcal{C}$.
\end{thm}

\begin{thm}[Implicit in \cite{grohe18}]\label{thm:nowhere-dense}
Let $\mathcal{C}$ be a nowhere dense class of structures.  Then \textsc{FO-Counting} is in FPT when restricted to structures from $\mathcal{C}$.
\end{thm}

We note that classes of graphs of bounded treewidth, bounded genus, and bounded degree are all nowhere dense.  For the monadic second-order case, we need stronger restrictions on the input to guarantee tractability.

\begin{thm}[\cite{arnborg91,courcelle-counting}]\label{thm:courcelle-count}
Let $\mathcal{C}$ be a class of structures of bounded treewidth.  Then \textsc{MSO-Counting} is in FPT when restricted to structures from $\mathcal{C}$.
\end{thm}

In this paper we are concerned with monotone classes.  Classes defined by a bound on the cliquewidth are not monotone, since deleting edges can increase the cliquewidth of a graph (every graph can be obtained by deleting edges from a complete graph, which has cliquewidth one).  However, classes of bounded treewidth, bounded genus and bounded degree are all closed under the deletion of both vertices an edges, and so are monotone.  More generally, if $\mathcal{C}$ is nowhere dense and every element $G'$ of $\mathcal{C}'$ is a subgraph of some element $G$ of $\mathcal{C}$, then $\mathcal{C}'$ is also nowhere dense.

In the special case of monotone classes, Theorem \ref{thm:nowhere-dense} is known to be the strongest possible result for \textsc{FO-Model-Checking} (and hence also for \textsc{FO-Counting}).

\begin{thm}[\cite{dvorak10,kreutzer11,grohe17}]\label{thm:somewhere-dense}
Let $\mathcal{C}$ be a class of structures that is closed under taking subgraphs.  Then, assuming FPT $\neq$ W[1], \textsc{FO-Model-Checking} is in FPT restricted to structures from $\mathcal{C}$ if and only if $\mathcal{C}$ is nowhere dense.
\end{thm}

\section{Properties of layered graph classes}
\label{sec:layers}

This section is concerned with the properties of graphs formed by combining layers which belong to specific graph classes.  

Given graph classes $\mathcal{C}_1$ and $\mathcal{C}_2$, we define the \emph{layered graph class} $\layer(\mathcal{C}_1,\mathcal{C}_2)$ to be the class of all graphs $G=(V,E)$ such that $V = V_1 \cup V_2$ and $E = E_1 \cup E_2$, where $G_1 = (V_1,E_1) \in \mathcal{C}_1$ and $G_2 = (V_2,E_2) \in \mathcal{C}_2$.  Note that the sets $V_1$ and $V_2$ (respectively $E_1$ and $E_2$) need not be disjoint: in particular, we will often be interested in the case in which $V_1 = V_2 = V$, so that we construct a layered graph by combining two graphs on the same vertex set.

We extend this notation in the obvious way to graphs involving more than two layers: given graph classes $\mathcal{C}_1,\ldots,\mathcal{C}_s$, we write $\layer(\mathcal{C}_1,\ldots,\mathcal{C}_s)$ for the layered graph class consisting of graphs $G = (V,E)$ where $V = V_1 \cup \cdots \cup V_s$, $E = E_1 \cup \cdots \cup E_s$, and $G_i = (V_i,E_i) \in \mathcal{C}_i$ for each $i \in \{1,\ldots,s\}$.

We observe that, when considering either graphs or structures, the $\layer$ operation is both associative and commutative.

Some graph properties are preserved under the layering operation: for example, if $s$ is any constant and classes $\mathcal{C}_1,\ldots,\mathcal{C}_s$ have bounded degree (respectively bounded vertex cover number), then so does $\layer(\mathcal{C}_1,\ldots,\mathcal{C}_s)$.  However, the same cannot be said for some more complex graph parameters: for example, if $\mathcal{T}$ is the class of acyclic graphs (which have treewidth one), then $\layer(\mathcal{T},\mathcal{T})$ contains all grids (as a grid can be obtained by combining two paths) and hence has unbounded treewidth.  In the remainder of this section we are concerned with identifying graphs that are guaranteed to belong to a layered graph class if we make specific assumptions about the individual layers; we will exploit these results when proving intractability in Section~\ref{sec:dichotomies}.

We begin with two simple results on single-layer graph classes which we will use in our arguments about layered graph classes. 

\begin{lma}\label{lma:find-deg1}
Let $\mathcal{C}$ be a monotone class of graphs of unbounded vertex cover number.  Suppose that $G$ has maximum degree one; then $G$ belongs to $\mathcal{C}$.
\end{lma}
\begin{proof}
Let $c$ be the number of connected components in $G$; it suffices to show that there is a graph $H \in \mathcal{C}$ which contains a matching on at least $c$ edges, as $G$ can then be obtained from $H$ by deleting edges and vertices.  Suppose there is no such $H$ in $\mathcal{C}$, so the largest matching in any element of $\mathcal{C}$ contains at most $c-1$ edges.  If $M$ is a maximal matching in any graph then the endpoints of edges in $M$ must form a vertex cover, so it follows that every element of $\mathcal{C}$ has vertex cover number at most $2c-2$, contradicting the fact that $\mathcal{C}$ has unbounded vertex cover number.
\end{proof}

\begin{lma}\label{lma:find-P3s}
Let $\mathcal{C}$ be a monotone class of graphs that is not at constant vertex cover distance from the class of graphs of maximum degree one.  Suppose that every connected component of $G$ is either an isolated vertex, a pair of vertices connected by an edge, or a three-vertex path.  Then $G$ belongs to $\mathcal{C}$.
\end{lma}
\begin{proof}
Set $c$ to be the number of connected components of $G$.  It suffices to show that there exists $H \in \mathcal{C}$ which contains at least $c$ vertex-disjoint copies of $P_3$ (the path on three vertices): a graph isomorphic to $G$ can be obtained from $H$ by deleting appropriate vertices and edges, and so by monotonicity such a graph must belong to $\mathcal{C}$.

Since $\mathcal{C}$ is not at constant vertex cover distance from the class of graphs of maximum degree one, we can find some $H \in \mathcal{C}$ such that, for any $X \subseteq V(H)$ with $|X| \leq 4c$, the maximum degree of $H$ is greater than one.  Fix such a graph $H$, and let $M$ be a maximal matching in $H$.  We will denote by $\widetilde{H}$ the graph obtained from $H$ by contracting every edge in $M$, and let $W$ denote the set of vertices in $\widetilde{H}$ obtained by contracting an edge in $M$.  Note that, by maximality of $M$, every edge in $\widetilde{H}$ has at least one endpoint in $M$.

We claim that $\widetilde{H}$ has vertex cover number at least $2c$.  If not, there exists a set $X \subset V(\widetilde{H})$ with $|X| \leq 2c$ such that $\widetilde{H} \setminus X$ contains no edges.  Setting $Y$ to be the set of vertices in $H$ corresponding to $X$ (so if $w \in W \cap X$, $Y$ contains both endpoints of the edge that was contracted to obtain $w$), we observe that $|Y| \leq 2|X| \leq 4c$.  Moreover, $E(H \setminus Y) \subseteq M$, since all edges not belonging to $M$ have been removed, so the maximum degree of $H \setminus Y$ is at most one.  This contradicts our choice of $H$, so we conclude that $\widetilde{H}$ has vertex cover number at least $2c$.

It follows that $\widetilde{H}$ contains a matching $\widetilde{M} = e_1,\ldots,e_c$.  Recall that every edge in $\widetilde{H}$ has at least one endpoint in $W$; for each $1 \leq i \leq c$ let $w_i$ be an endpoint of $e_i$ which belongs to $W$ (note that $w_i \neq w_j$ for $i \neq j$).  Suppose that each $w_i$ was obtained by contracting the edge $u_iv_i$.  For each $1 \leq i \leq c$, there exists $z_i \in V(H)$ which is adjacent to either $u_i$ or $v_i$ ($z_i$ is either the other endpoint of $e_i$ in $\widetilde{H}$ or, if this other endpoint belongs to $W$, one of the corresponding vertices in $H$); we will assume without loss of generality that $z_i$ is adjacent to $v_i$ in each case.  Thus, for each $i$, $u_iv_iz_i$ is a path on three vertices.  It remains to argue that these copies of $P_3$ are disjoint.  Suppose, for a contradiction, that $x \in \{u_i,v_i,z_i\} \cap \{u_j,v_j,z_j\}$ for $i \neq j$.  If $x \in \{u_i,v_i\} \cap \{u_j,v_j\}$ then, as we know that $w_i \neq w_j$ for $i \neq j$, we would have that two edges of $M$ share a vertex, contradicting the fact that $M$ is a matching.  If $x = z_i \in \{u_j,v_j\}$ (or, symmetrically, $x = z_j \in \{u_i,v_i\}$) then we have a contradiction to the fact that both $M$ and $\widetilde{M}$ are matchings, as $z_i$ either corresponds to $w_j$ (meaning that $\widetilde{M}$ is not a matching) or $z_i$ and $w_j$ are obtained by contracting edges with a common endpoint (meaning that $M$ is not a matching).  Similarly, if $x = z_i = z_j$, at least one of $M$ and $\widetilde{M}$ is not a matching.

Thus we see that $H$ contains at least $c$ vertex-disjoint copies of $P_3$, as required.
\end{proof}

Using Lemma \ref{lma:find-deg1}, we now give a sufficient condition for a class formed from three layers to contain walls of arbitrary height.

\begin{lma}\label{lma:3unbddVC}
Let $\mathcal{C}_1, \mathcal{C}_2$ and $\mathcal{C}_3$ be monotone classes of graphs of unbounded vertex cover number.  Then, for any $h \in \mathbb{N}$, there is an element of $\layer(\mathcal{C}_1,\mathcal{C}_2,\mathcal{C}_3)$ which is isomorphic to $W_h$.
\end{lma}
\begin{proof}
Notice that $W_h$ is a bipartite graph with maximum degree three and hence is 3-edge-colourable.  Therefore we can partition the edges of $W_h = (V,E)$ into three sets $E_1$, $E_2$ and $E_3$ such that each set $E_i$ consists of vertex-disjoint edges.  It follows that each graph $(V,E_i)$ has maximum degree one and hence, as each class $\mathcal{C}_i$ is monotone and has unbounded vertex cover number, we know from Lemma \ref{lma:find-deg1} that there is some graph $G_i \in \mathcal{C}_i$ which is isomorphic to $(V,E_i)$.  It follows that $W_h = \layer(G_1,G_2,G_3) \in \layer(\mathcal{C}_1,\mathcal{C}_2,\mathcal{C}_3)$, as required.
\end{proof}

Using both Lemmas \ref{lma:find-deg1} and \ref{lma:find-P3s}, we can also give a sufficient condition for a two-layer graph class to contain \emph{subdivisions} of walls of arbitrary height.

\begin{lma}\label{lma:notconstdel}
Let $\mathcal{C}_1$ and $\mathcal{C}_2$ be monotone classes of graphs of unbounded vertex cover number, and suppose further that $\mathcal{C}_1$ is not at constant vertex deletion distance from the class of graphs of maximum degree one.  Then, for any $h \in \mathbb{N}$, $\layer(\mathcal{C}_1,\mathcal{C}_2)$ contains a subdivision of $W_h$ in which each horizontal edge is subdivided once and each vertical edge is subdivided twice.
\end{lma}
\begin{proof}
Let $W_h'$ denote the graph obtained from $W_h$ by subdividing every horizontal edge three times and every vertical edge twice.  We will show that there exist graphs $G_1$ and $G_2$ such that $W_h' = \layer(G_1,G_2)$, where $G_2$ has maximum degree one and every connected component of $G_1$ is either an isolated vertex, two vertices connected by an edge, or a path on three vertices.  Applying Lemmas \ref{lma:find-deg1} and \ref{lma:find-P3s}, we see that $G_1 \in \mathcal{C}_1$ and $G_2 \in \mathcal{C}_2$ and hence $W_h' \in \layer(\mathcal{C}_1,\mathcal{C}_2)$, as required.

To show that there exist suitable graphs $G_1$ and $G_2$, it suffices to partition the edges of $W_h'$ into two sets $E_1$ and $E_2$ such that $E_2$ is an independent set of edges and every connected component in the graph $(V(W_h),E_1)$ is either an isolated vertex, a single edge or a path on three vertices; a suitable partition of the edges of $W_3'$ is illustrated in Figure \ref{fig:evilgenius}.  

\begin{figure}
\centering
\includegraphics[width = 0.4 \linewidth]{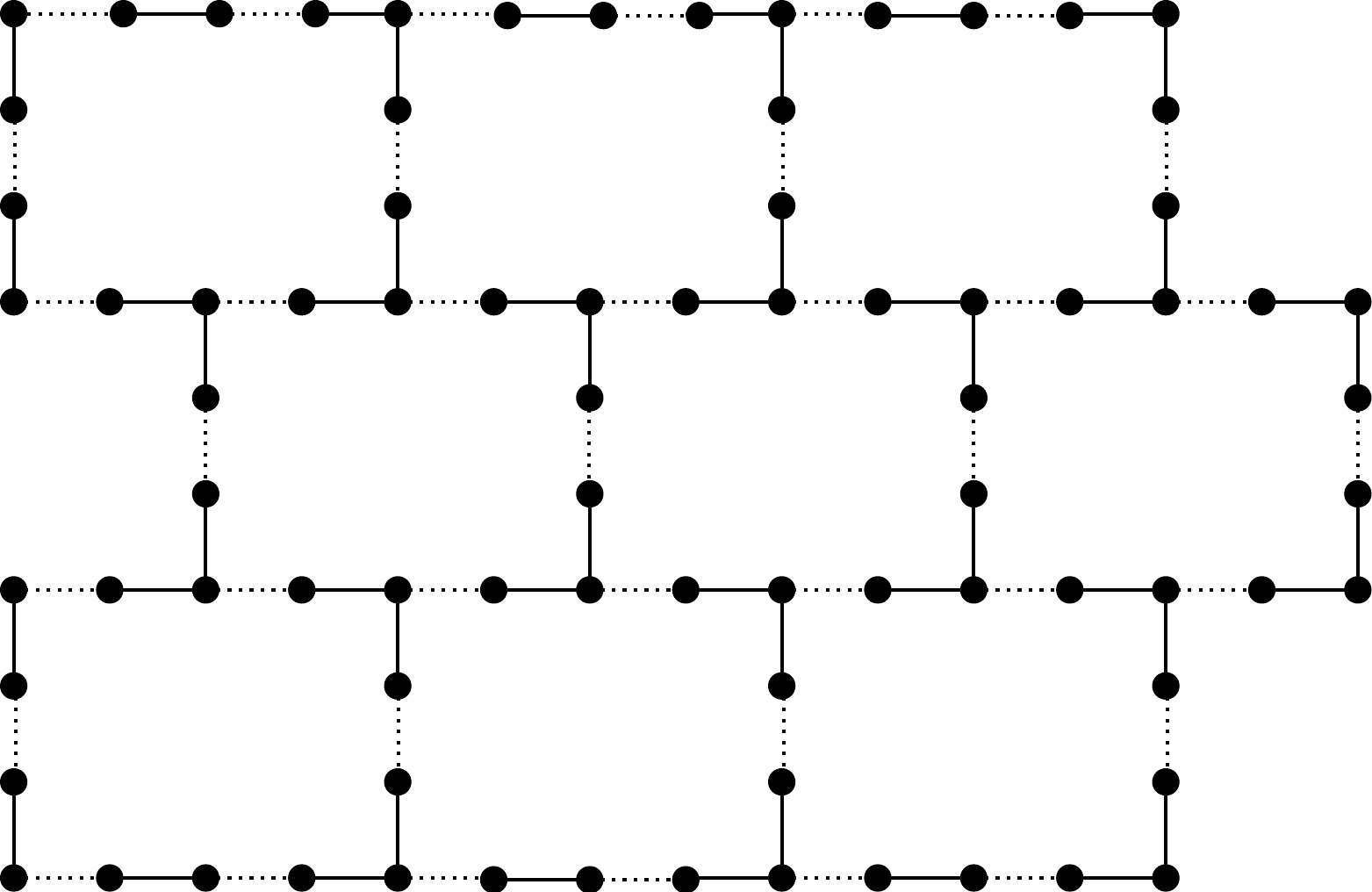}
\caption{The edges of a subdivided wall of height three can be partitioned into two sets, where the first (dotted) corresponds to a collection of vertex-disjoint edges, and the second (solid) corresponds to a collection of vertex-disjoint edges and copies of $P_3$.}
\label{fig:evilgenius}
\end{figure}

We first consider those edges obtained by subdividing vertical edges of $W_h$.  Set $E_2'$ to be the set of edges in $W_h'$ that are obtained from subdividing vertical edges and are not incident with any branch vertex (i.e. the middle edge in each $P_4$ that replaces a vertical edge in $W_h$).  Note that every connected component $C_i$ in $W_h' \setminus E_2'$ consists of a path $P_i$ (formed by the subdivided horizontal edges) together with some pendant edges (from subdivided vertical edges); moreover, no two pendant edges are incident with the same vertex on the path and no two vertices of degree three (those vertices on the path incident with pendant edges) are adjacent.  For each $1 \leq i \leq h + 1$, we assign exactly half of the edges of $P_i$ to the set $E_2''$: we assign to $E_2''$ the edge incident with $(i,1)$ and every second edge along the path.  Setting $E_2 := E_2' \cup E_2''$, we note that $E_2$ is a set of independent edges, as required.  

It remains to show that $E_1 := E(W_h' \setminus E_2)$ has the required properties.  Let $F$ be a connected component of $(V(W_h'),E_1)$.  Note first that $F$ is contained in some component $C_i$ of $W_h' \setminus E_2'$.  We further observe that $F$ contains at most one edge from $P_i$, since $C_i$ is a tree and no component of $P_i \setminus E_2''$ contains more than one edge.  Since no two edges pendant to $P_i$ in $C_i$ are incident with the same vertex or with adjacent vertices, it follows that $F$ contains at most one pendant edge.  Therefore $F$ contains at most two edges and is either an isolated vertex, a single edge or a path on three vertices.  This completes the proof.
\end{proof}

\section{Dichotomy results}
\label{sec:dichotomies}

In this section we prove our dichotomy results for both first-order and monadic second-order model checking, before discussing the explicit consequences for graphs encoded with an incidence relation in Section \ref{sec:graphs}.  Specifically, we prove the following results.

\begin{thm}\label{thm:FO-dichotomy}
Let $\mathcal{C}_1,\ldots,\mathcal{C}_s$ be monotone classes of relational structures such that, for each $1 \leq i \leq s$, \textsc{FO-Model-Checking} (respectively \textsc{FO-Counting}) is in FPT when restricted to structures from $\mathcal{C}_i$.  Then, assuming FPT $\neq$ W[1], \textsc{FO-Model-Checking} (respectively \textsc{FO-Counting}) is in FPT restricted to structures from $\layer(\mathcal{C}_1,\ldots,\mathcal{C}_s)$ if and only if either
\begin{enumerate}
\item there is some $j \in [s]$ such that, for all $i \in [s]\setminus \{j\}$, $\mathcal{C}_i$ has bounded vertex cover number, or
\item each of $\mathcal{C}_1,\ldots,\mathcal{C}_s$ is at constant vertex cover distance from the class of bounded degree graphs.
\end{enumerate}
\end{thm}

\begin{thm}\label{thm:MSO-dichotomy}
Let $\mathcal{C}_1,\ldots,\mathcal{C}_s$ be monotone classes of relational structures such that, for each $1 \leq i \leq s$, \textsc{MSO-Model-Checking} (respectively \textsc{MSO-Counting}) is in FPT when restricted to structures from $\mathcal{C}_i$.  \textsc{MSO-Model-Checking} (respectively \textsc{MSO-Counting}) is in FPT restricted to structures from $\layer(\mathcal{C}_1,\ldots,\mathcal{C}_s)$ if either
\begin{enumerate}
\item there is some $j \in [s]$ such that, for all $i \in [s]\setminus \{j\}$, $\mathcal{C}_i$ has bounded vertex cover number, or
\item there exist $i,j \in [s]$ such that $\mathcal{C}_i$ and $\mathcal{C}_j$ are both at constant vertex cover distance from the class of graphs of maximum degree one and, for all $\ell \in [s] \setminus \{i,j\}$, $\mathcal{C}_{\ell}$ has bounded vertex cover number.
\end{enumerate}
Otherwise, unless \textsc{SAT} can be solved in subexponential time $2^{o(n)}$, \textsc{MSO-Model-Checking} is not in XP.
\end{thm}

The positive direction of Theorem \ref{thm:FO-dichotomy}, and the positive direction of Case 2 of Theorem \ref{thm:MSO-dichotomy} can be deduced by considering only the structural properties of the resulting graph classes.  However, for simplicity, we use the same proof strategy for the positive direction of both theorems; this relies on the following lemma, which involves a fairly standard application of the \emph{interpretation method}.

\begin{lma}\label{lma:splitVC}
Fix $\Phi$ to be a class of formulas such that $\FO \subseteq \Phi$, and let $\tau$ be a vocabulary.  Let $\mathcal{C}$ be a class of structures on which  \textsc{$\Phi$-Model-Checking} (respectively \textsc{$\Phi$-Counting}) can be solved in $f(k) \cdot n^{\mathcal{O}(1)}$ for some computable, non-decreasing function $f$, where $n$ is the size of the structure and $k$ is the size of the formula.  Suppose that $\mathcal{C}'$ is a class of structures at vertex cover distance at most $c$ from $\mathcal{C}$ and that, for any $\mathfrak{A} \in \mathcal{C}'$, we can find, in time polynomial in the size of $\mathfrak{A}$, a subset $X$ of the universe which witnesses this vertex cover distance.  Then there is a constant $\alpha$ such that \textsc{$\Phi$-Model-Checking} (respectively \textsc{$\Phi$-Counting}), restricted to structures from $\mathcal{C}'$, can be solved in time $f(\alpha r_{\max}^{c+1}k^2) \cdot n^{\mathcal{O}(1)}$, where $r_{\max}$ is the maximum arity of any relation $R \in \tau$.
\end{lma}
\begin{proof}
Let $(\mathfrak{A},\phi)$ be the input to an instance of \textsc{$\Phi$-Model-Checking} (respectively \textsc{$\Phi$-Counting}), where $\mathfrak{A} \in \mathcal{C}'$.  We will denote by $A$ and $G$ respectively the universe and Gaifman graph of $\mathfrak{A}$.  By assumption, we can find in polynomial time a set $X = \{x_1,\ldots,x_c\} \subseteq A$ such that $G \setminus X$ belongs to $\mathcal{C}$.

Our strategy is to define a new vocabulary $\tau'$, a $\tau'$-structure $\mathfrak{A}'$ (with the same universe $A$) which belongs to $\mathcal{C}$, and a formula $\phi'$ such that $\phi'(\mathfrak{A}') = \phi(\mathfrak{A})$.  Provided that this construction can be carried out efficiently and that neither the size of the structure nor the size of the formula increases too much, the result will follow immediately.

We begin by defining the new vocabulary $\tau'$.  We first introduce a unary relation $R_x$ for each $x \in X$; this will be used to identify the element of the universe $A$ that is equal to $x$.  Now fix a relation $R \in \tau$, and suppose that $R$ has arity $r > 1$.  We introduce a collection of relations $\tau_R$ which will be used to encode tuples which can be extended with appropriate elements of $X$ to give tuples belonging to $R$.  The set $\tau_R$ contains, for every injective function $\pi$ from a subset of $[r]$ to $X$, a relation $R_{\pi}$ whose arity is equal to $r - |\dom(\pi)|$, where $\dom(\pi)$ denotes the domain of $\pi$.  If $\tau_{>1}$ denotes the set of relations in $\tau$ whose arity is greater than one, we set
$$\tau' = \tau \cup \bigcup_{R \in \tau_{>1}} \tau_R \cup \bigcup_{x \in X} R_x.$$

We now define our $\tau'$-structure $\mathfrak{A}'$, which has universe $A \setminus X$.  The interpretation of each relation symbol $Q \in \tau'$ is defined as follows:
\begin{itemize}
\item If $Q \in \tau$ and $Q$ has arity one, then $Q^{\mathfrak{A}'} = Q^{\mathfrak{A}}$.
\item If $Q \in \tau$ and $Q$ has arity $r > 1$, then
$$Q^{\mathfrak{A}'} = \{(a_1,\ldots,a_r) \in Q^{\mathfrak{A}}: a_1,\ldots,a_r \notin X\}.$$
\item If $Q \notin \tau$ and $Q = R_x$ for some $x \in X$, then $R_x^{\mathfrak{A}'} = \{(x)\}$.
\item If $Q \notin \tau$ has arity $s$ and $Q = R_{\pi} \in \tau_R$ for some relation $R \in \tau$ of arity $r$, then, for $a_1,\ldots,a_s \in A \setminus X$, we have $(a_1,\ldots,a_s) \in Q^{\mathfrak{A}'}$ if and only if $(b_1,\ldots,b_r) \in Q^{\mathfrak{A}}$ where, if $[r] \setminus \dom(\pi) = \{j_1,\ldots,j_s\}$ with $j_1 < \cdots < j_s$,
\begin{equation*}
b_i = \begin{cases}
			\pi(i)		& \text{if $i \in dom(\pi)$} \\
			a_{\ell} 	& \text{if $i = j_{\ell}$.}
	  \end{cases}
\end{equation*}
\end{itemize} 
We claim that $\mathfrak{A}'$ belongs to $\mathcal{C}$.  To see this, observe that the Gaifman graph $G'$ of $\mathfrak{A}$ is a subgraph of $G$, and moreover that (as the only relations involving elements of $X$ are unary) $G'$ does not contain any edge incident with $X$.  Thus, as we know that $G \setminus E(X,*)$ belongs to $\mathcal{C}$ by choice of $X$, and $\mathcal{C}$ is closed under deletion of vertices and/or edges, it follows that $G' \in \mathcal{C}$.  Note further that $\mathfrak{A}'$ can clearly be constructed from $\mathfrak{A}$ in time polynomial in the size of $\mathfrak{A}$ (and also that the size of $\mathfrak{A}'$ is bounded by a polynomial function of the size of $\mathfrak{A}$).

It remains only to construct a formula $\phi'$ such that $\phi'(\mathfrak{A}') = \phi(\mathfrak{A})$ and $|\langle \phi' \rangle| = \mathcal{O}\left(r_{\max}^{c+1}|\phi|^2\right)$.  To do this it suffices to show that, for any relation $R \in \tau$ of arity $r > 1$, we can construct a formula $\theta_R \in \FO[\tau'] \subseteq \Phi[\tau']$, whose size is not too large, such that $\mathfrak{A}' \models \theta_R(a_1,\ldots,a_r)$ if and only if $(a_1,\ldots,a_r) \in R^{\mathfrak{A}}$.  It is straightforward to verify that the following formula has this property, where for any tuple $(a_1,\ldots,a_r)$ we write $(a_1,\ldots,a_r)|_Y$ for the tuple obtained by removing all elements that do not belong to $Y$:
\begin{align*}
\theta_R(a_1,\ldots,a_r) := & \\
		R^{\mathfrak{A}'}(a_1,\ldots,a_r) & \vee \bigvee_{R_{\pi} \in \tau_R} \left(R^{\mathfrak{A}'}_{\pi} \left((a_1,\ldots,a_r)|_{[r] \setminus \dom(\pi)}\right)  \wedge \bigwedge_{i \in \dom(\pi)} R_{\pi(i)}(a_i)\right).
\end{align*}
To complete the proof, we observe that the length of $\theta_R$ is $\mathcal{O}(|\tau_R|\cdot r) = \mathcal{O}(r^{c+2}) = \mathcal{O}(r_{\max}^{c+2})$.  Since $\theta_R$ will replace an occurrence of $R$ in the formula $\phi$, where $\arity(R) = r$, it follows that replacing all such occurrences in this way will increase the length of the formula by a factor of at most $\mathcal{O}(r_{\max}^{c+1})$.  Thus there is an encoding $\langle \phi' \rangle$ of $\phi'$ with
\begin{align*}
|\langle \phi' \rangle| & = \mathcal{O}\left(r_{\max}^{c+1}|\phi| \cdot \log(r_{\max}^{c+1}|\phi|)\right) \\
		& = \mathcal{O}\left(r_{\max}^{c+1}|\phi| \left((c+1)\log r_{\max} + \log |\phi|\right)\right) \\
		& = \mathcal{O}\left(r_{\max}^{c+1}|\phi|^2\right),
\end{align*}
as required.
\end{proof}

An easy corollary of this result proves part of the positive direction of both Theorems \ref{thm:FO-dichotomy} and~\ref{thm:MSO-dichotomy}.

\begin{cor}\label{cor:tractable-VC}
Fix $\Phi$ to be a class of formulas such that $\FO \subseteq \Phi$.  Let $\mathcal{C}_1,\ldots,\mathcal{C}_{s-1}$ be classes of relational structures with bounded vertex cover number, and let $\mathcal{C}_s$ be a monotone class of relational structures on which \textsc{$\Phi$-Model-Checking} (respectively \textsc{$\Phi$-Counting} is in FPT.  Then \textsc{$\Phi$-Model-Checking} (respectively \textsc{$\Phi$-Counting}) is in FPT restricted to structures from $\layer(\mathcal{C}_1,\ldots,\mathcal{C}_s)$.
\end{cor}
\begin{proof}
By definition of bounded vertex cover number, for each $1 \leq i \leq s-1$, there is a constant $c_i$ such that the vertex cover number of every structure in $\mathcal{C}_i$ is at most $c_i$; set $c = \sum_{1 \leq i \leq s-1} c_i$.  It follows from this, together with the fact that $\mathcal{C}_s$ is monotone, that every element of $\layer(\mathcal{C}_1,\ldots,\mathcal{C}_s)$ is at vertex cover distance at most $c$ from $\mathcal{C}_s$: given any $\mathfrak{A} = \layer(\mathfrak{A}_1,\ldots,\mathfrak{A}_s)$ with $\mathfrak{A}_i \in \mathcal{C}_i$, it suffices to delete all edges incident with $\bigcup_{1 \leq i \leq s-1} X_i$ where each $X_i$ is a vertex cover for $A_i$.  Moreover, we can find a suitable $X_i$ in polynomial time, as we are seeking a vertex cover of constant size.  The result now follows immediately from Lemma \ref{lma:splitVC}.
\end{proof}

To complete the proof of the positive direction of Theorem \ref{thm:FO-dichotomy}, we need to consider the case in which all classes are at constant vertex cover distance from a class of graphs of bounded maximum degree.

\begin{cor}\label{lma:FO-tractable-ABD}
Fix a constant $d \in \mathbb{N}$ and let $\mathcal{C}_1,\ldots,\mathcal{C}_s$ be classes of relational structures which are at constant vertex cover distance from the class of graphs of maximum degree at most $d$.  Then \textsc{FO-Model-Checking} (respectively \textsc{FO-Counting}) is in FPT restricted to structures from $\layer(\mathcal{C}_1,\ldots,\mathcal{C}_s)$.
\end{cor}
\begin{proof}
Suppose that each class $\mathcal{C}_1,\ldots,\mathcal{C}_s$ is at vertex cover  distance at most $c$ from the class of graphs of maximum degree at most $d$.  Fix an arbitrary element $\layer(\mathfrak{A}_1,\ldots,\mathfrak{A}_s) \in \layer(\mathcal{C}_1,\ldots,\mathcal{C}_s)$, and let $G_i$ be the Gaifman graph of $\mathfrak{A}_i$ for each $i$.  The fact that $G_i$ is at vertex cover distance at most $c$ from the class of maximum degree $d$ implies that $G_i$ contains at most $c + c^2$ vertices of degree greater than $d$; thus we can certainly find in polynomial time a set $X_i$, of cardinality at most $c + c^2$, such that $G_i \setminus E_{G_i}(X_i,*)$ has maximum degree at most $d$.  Setting $X = \bigcup_{1 \leq i \leq s} X_i$, we see that $|X| \leq s(c+c^2)$ and that, if $G$ is the Gaifman graph of $\layer(\mathfrak{A}_1,\ldots,\mathfrak{A}_s)$, then $G \setminus E_G(X,*)$ has maximum degree at most $sd$.  Recalling from Theorem \ref{thm:nowhere-dense} that \textsc{FO-Model-Checking} and \textsc{FO-Counting} are in FPT when restricted to the class of graphs of bounded maximum degree, the result now follows immediately from Lemma \ref{lma:splitVC}.
\end{proof}

For the negative direction of Theorem \ref{thm:FO-dichotomy}, we recall that the subgraph isomorphism problem is a special case of \textsc{FO-Model-Checking}, so it suffices to prove intractability for this specific problem.  The subgraph isomorphism problem can be defined formally in terms of embeddings: an embedding of a graph $H$ into a graph $G$ is an injective mapping $\theta$ from $V(H)$ to $V(G)$ such that, whenever $uv$ is an edge in $H$, we have that $\theta(u)\theta(v)$ is an edge in $G$.  The problem \emb\, is then defined as follows.
\begin{framed}
\noindent
\emb\\
\textit{Input:} Two graphs $G$ and $H$.\\
\textit{Parameter:} $k = |H|$.\\
\textit{Question:} Is there an embedding of $H$ into $G$?
\end{framed}
The corresponding counting problem, \embcount, asks for the number of embeddings of $H$ into $G$.  We refer to $G$ as the \emph{host graph} and $H$ as the \emph{pattern graph}.  In general, \emb\, is W[1]-complete \cite{downey1995fixed} and \embcount\, is \#W[1]-complete \cite{flum04}.

\begin{lma}\label{lma:starfor+matching}
Let $\mathcal{C}_1$ be the class of star forests and $\mathcal{C}_2$ the class of graphs with maximum degree one.  Then \emb~is W[1]-hard even if the host graph is restricted to $\mathcal{C} = \layer(\mathcal{C}_1,\mathcal{C}_2)$.
\end{lma}
\begin{proof}
We give a reduction from the following problem, shown to be W[1]-hard in \cite{downey1995fixed}.

\begin{framed}
\noindent
\textbf{$p$}-\textsc{Clique}\newline
\textit{Input: }A graph G and $k \in \mathbb{N}$\newline
\textit{Parameter: }$k$\newline
\textit{Question: } Is there a $k$-clique in G?
\end{framed}

Let $(G,k)$ be the input to an instance of \textbf{$p$}-\textsc{Clique}, where $G$ has $n$ vertices and $m$ edges; we assume without loss of generality that $k>3$. We first construct a pair of graphs $\widetilde{G}$ and $\widetilde{H}$ such that there is an embedding of $\widetilde{H}$ into $\widetilde{G}$ if and only if $G$ contains a $k$-clique. We then show that the edges of $\widetilde{G}$ can be partitioned into two sets $\widetilde{E}_1$ and $\widetilde{E}_2$ such that $\widetilde{G}_1=(V(\widetilde{G}),\widetilde{E}_1)$ is a star forest and $\widetilde{G}_2=(V(\widetilde{G}),\widetilde{E}_2)$ has maximum degree one.

To obtain $\widetilde{G}$ from $G$ we simply subdivide every edge of $G$ twice. The graph $\widetilde{H}$ is obtained from the $k$-clique in the same way. Notice that $\widetilde{H}$ has precisely $k + 2 \binom{k}{2}$ vertices, and the number of vertices in $\widetilde{G}$ is $n + 2m$. 

It is easy to see that $\widetilde{G}$ contains an embedding of $\widetilde{H}$ whenever $G$ contains a $k$-clique.
Conversely, suppose that $\widetilde{G}$ contains an embedding $\theta$ of $\widetilde{H}$. We refer to vertices of $\widetilde{G}$ and $\widetilde{H}$ that subdivide edges as subdividing vertices, and to the remaining vertices as branch vertices.  

Let $B$ denote the set of $k$ branch vertices in $\widetilde{H}$, and set $U = \theta(B)$ so that $U$ is the set of vertices in $\widetilde{G}$ to which elements of $B$ are mapped by the embedding.  Note that $|U| = |B| = k$.

Since $k>3$, every vertex in the $k$-clique, and hence every branch vertex in $\widetilde{H}$, has degree at least three. It follows that all elements of $U$ must be branch vertices in $\widetilde{G}$, since all subdividing vertices in $\widetilde{G}$ have degree exactly two.

We claim that $U$ forms a clique in $G$.  To see that this is true, observe that the number of edges on the shortest path between two branch vertices $v$ and $w$ in $\widetilde{G}$ is precisely three times the number of edges on the shortest path between $v$ and $w$ in $G$.  For any $u_i \neq u_j \in U$, there is a three-edge path between the corresponding vertices in $\widetilde{H}$ and hence, by the definition of an embedding, there is also a three-edge path from $u_i$ to $u_j$ in $\widetilde{G}$.  It follows that there must be a one-edge path between $u_i$ and $u_j$ in $G$, in other words that $u_i$ and $u_j$ are adjacent in $G$.  Since this holds for all $u_i \neq u_j \in U$, we conclude that $U$ does indeed induce a $k$-clique in $G$.

Finally, it remains to show $\widetilde{G}$ can be partitioned into graphs $\widetilde{G_1}$ and $\widetilde{G_2}$ such that $\widetilde{G}_1=(\widetilde{V}_1,\widetilde{E}_1)$ is a star forest and $\widetilde{G}_2=(\widetilde{V}_2,\widetilde{E}_2)$ has maximum degree one. Let $\widetilde{E}_1$ be the set of edges in $\widetilde{G}$ which are incident to branch vertices, and $\widetilde{E}_2$ the remaining edges from $\widetilde{E}$. To see that $\widetilde{G}_1$ is a star forest, observe that (a) all subdividing vertices have degree one in $\widetilde{G_1}$, and (b) each connected component of $\widetilde{G_1}$ contains exactly one branch vertex.  For $\widetilde{G_2}$, note that the only edges in $\widetilde{E_2}$ are those not incident to any branch vertex.  Since each edge of $G$ was subdivided exactly once, it follows that $\widetilde{E}_2$ is a set of independent edges in $\widetilde{G}$ and hence that $\widetilde{G_2}$ has maximum degree one.
\end{proof}

As an aside, we note that the reduction given in Lemma \ref{lma:starfor+matching} is parsimonious, so the following result for the counting version of the problem follows immediately.

\begin{cor}
Let $\mathcal{C}_1$ and $\mathcal{C}_2$ be isomorphism-closed monotone graph classes of unbounded vertex cover number, and suppose further that $\mathcal{C}_1$ is not at constant vertex cover distance from the class of graphs of bounded maximum degree.  Then \embcount~is W[1]-hard when restricted to host graphs from $\mathcal{C} = \layer(\mathcal{C}_1,\mathcal{C}_2)$.
\end{cor}

To complete the proof of Theorem \ref{thm:FO-dichotomy} it suffices to show that, if our graph classes do not meet either of the conditions of the theorem, then one class contains all star forests and another contains all graphs of maximum degree one.

\begin{proof}[Proof of Theorem \ref{thm:FO-dichotomy}]
The positive direction is immediate from Corollaries \ref{cor:tractable-VC} and \ref{lma:FO-tractable-ABD}; by Lemma \ref{lma:starfor+matching} it suffices for the reverse direction to demonstrate that, if at least two classes have unbounded vertex cover number and at least one is not at constant vertex cover distance from a class of graphs of bounded degree, then one class contains all finite star forests and another contains all finite graphs of maximum degree one.

We may assume without loss of generality that $\mathcal{C}_1$ is a class of graphs that is not at constant vertex cover distance from any class of graphs of bounded degree, and that $\mathcal{C}_2$ is a class of graphs of unbounded vertex cover number.  We know from Lemma \ref{lma:find-deg1} that $\mathcal{C}_2$ contains all finite graphs of maximum degree one, so it remains only to show that $\mathcal{C}_1$ contains all finite star forests.

Let $F$ be an arbitrary star forest; we will argue that $F \in \mathcal{C}_1$.  Let $\Delta$ be the maximum degree of $F$, and suppose that $F$ has exactly $c$ connected components.  We will show that $\mathcal{C}_1$ contains $c K_{1,\Delta}$, the star forest consisting of $c$ identical connected components, each isomorphic to $K_{1,\Delta}$; the fact that $F \in \mathcal{C}_1$ will then follow immediately from monotonicity of $\mathcal{C}_1$.  Since $\mathcal{C}_1$ is not at constant vertex cover distance from a graph of bounded degree, there must be some graph $G \in \mathcal{C}_1$ which has at least $c(\Delta + 1)$ vertices of degree at least $c(\Delta + 1)$.  In $G$ we find a collection of $c$ vertex-disjoint copies of $K_{1,\Delta}$ greedily as follows: pick any vertex of degree at least $\Delta$ and delete it together with $\Delta$ of its neighbours.  The deleted vertex set induces a graph which contains $K_{1,\Delta}$ as a subgraph, while the degree of any vertex in the rest of $G_1$ decreases by at most $\Delta + 1$.  Thus, we will be able to repeat this process $c$ times to obtain our disjoint copies of $K_{1,\Delta}$, as required.
\end{proof}

We now turn our attention to the case of monadic second-order logic.  One part of the positive direction is immediate from Corollary \ref{cor:tractable-VC}; for the second part we adapt slightly the argument in Corollary \ref{lma:FO-tractable-ABD}.

\begin{cor}\label{lma:MSO-tractable-deletion}
Let $\mathcal{C}_1$ and $\mathcal{C}_2$ be classes of relational structures which are at constant vertex cover distance from the class of structures of maximum degree one.  Then \textsc{MSO-Model-Checking} (respectively \textsc{MSO-Counting}) is in FPT restricted to structures from $\layer(\mathcal{C}_1,\mathcal{C}_2)$.
\end{cor}
\begin{proof}
Let $\mathfrak{A}_1 \in \mathcal{C}_1$ and $\mathfrak{A}_2 \in \mathcal{C}_2$ be relational structures with Gaifman graphs $G_1$ and $G_2$ respectively.  Following the same reasoning as in the proof of Lemma \ref{lma:FO-tractable-ABD}, we can find in polynomial time constant-sized sets $X_1 \subseteq V(G_1)$ and $X_2 \subseteq V(G_2)$ such that $G_1 \setminus E_{G_1}(X_1,*)$ and $G_2 \setminus E_{G_2}(X_2,*)$ both have maximum degree at most one.  Note that $\layer(G_1,G_2) \setminus E_{\layer(G_1,G_2)}(X_1 \cup X_2,*)$ is isomorphic to a subgraph of $\layer(G_1 \setminus E_{G_1}(X_1,*), G_2 \setminus E_{G_2}(X_2,*))$ and so has maximum degree at most two; since a graph with maximum degree at most two is a disjoint union of paths and cycles, it follows that $\layer(G_1,G_2)$ and hence $\layer(\mathfrak{A}_1,\mathfrak{A}_2)$ is at constant vertex cover distance from the class of graphs of bounded treewidth.  Moreover, we can certainly find a witnessing deletion in polynomial time (it suffices to delete all vertices of degree greater than two).  Thus the result follows immediately from Lemma \ref{lma:splitVC} together with Theorem \ref{thm:courcelle-count}.
\end{proof}

For the reverse direction, we will rely on a partial converse to Courcelle's Theorem (the decision version of Theorem \ref{thm:courcelle-count}) due to Kreutzer.  Before stating this result, we need a definition.

\begin{adef}[\cite{kreutzer10}]
The treewidth of a class $\mathcal{C}$ of graphs is \textbf{strongly unbounded} by a function $f \colon \mathbb{N} \rightarrow \mathbb{N}$ if there is $\varepsilon < 1$ and a polynomial $p(x)$ such that, for all $n \in \mathbb{N}$, there is a graph $G_n \in \mathcal{C}$ with the following properties:
\begin{enumerate}
\item the treewidth of $G_n$ is between $n$ and $p(n)$ and is not
bounded by $f(|G_n|)$, and
\item given $n$, $G_n$ can be constructed in time $2^{n^\varepsilon}$.
\end{enumerate}
The degree of the polynomial $p$ is called the gap-degree of $\mathcal{C}$ (with respect to $f$). 
\end{adef}

We can now state the version of the theorem that we will use.

\begin{thm}[\cite{kreutzer10}]\label{thm:KreutzerHard}
Let $\mathcal{C}$ be a class of graphs closed under subgraphs.  If the treewidth of $\mathcal{C}$ is strongly unbounded by $\log^{28\gamma} n$, where $\gamma > 1$ is larger than the gap-degree of $\mathcal{C}$, then MSO-Model-Checking is not in XP unless \textsc{SAT} can be solved in subexponential time $2^{o(n)}$.
\end{thm}

Using this result, we show that \textsc{MSO-Model-Checking} is unlikely to be in FPT when restricted to any class that contains arbitrarily large walls whose edges are subdivided at most a constant number of times.  We write $\tw(G)$ for the treewidth of a graph $G$.

\begin{lma}\label{lma:walls-hard}
Fix constants $c_1$ and $c_2$ and let $\mathcal{C}$ be a monotone class of graphs which, for each $h \in \mathbb{N}$, contains a subdivision of $W_h$ in which each horizontal edge is subdivided exactly $c_1$ times and each vertical edge is subdivided exactly $c_2$ times.  Then, unless \textsc{SAT} can be solved in subexponential time $2^{o(n)}$, \textsc{MSO-Model-Checking} is not in XP when restricted to graphs from $\mathcal{C}$.
\end{lma}
\begin{proof}
We will denote by $W_h^{[c_1,c_2]}$ the graph obtained from $W_h$ by subdividing each horizontal edge exactly $c_1$ times and each vertical edge exactly $c_2$ times.

By Theorem \ref{thm:KreutzerHard}, it suffices to show that we can fix $\gamma > 1$ such that the treewidth of $\mathcal{C}$ is strongly unbounded by $\log^{28\gamma}n$ and $\gamma$ is larger than the gap-degree of $\mathcal{C}$.  To demonstrate this, we set
\begin{equation*}
G_n = \begin{cases}
			W_n^{[c_1,c_2]}		& \text{if $n > N_0$},\\
			W_{N_0}^{[c_1,c_2]}	& \text{otherwise,}
	  \end{cases}
\end{equation*}
where $N_0$ is a constant to be determined later; we must choose $N_0$ large enough that $\tw(G_n)$ is not bounded by $\log^{28 \gamma}n$.  Note that we can construct $G_n$ in time polynomial in $n$ (so certainly we can do this in time $2^{n^{\epsilon}}$).  Moreover, for every $n$, we have $n \leq \tw(G_n) \leq N_0 n$ for all $n$, so the gap-degree of $\mathcal{\mathcal{C}}$ is one.  We can therefore fix any $\gamma > 1$; for convenience we set $\gamma = 29/28$.  To complete the proof it suffices to show that we can choose a constant $N_0$ so that $\tw(G_n) > (\log |G_n|)^{29}$ for all $n$.

Observe that $W_n$ has $2n^2 + 4n$ vertices, $2n^2 + 3n - 1$ horizontal edges, and $n^2 + n$ vertical edges.  Therefore 
\[|W_n^{[c_1,c_2]}| = (2c_1 + c_2 + 2)n^2 + (3c_1 + c_2 + 4)n - c_1 \leq (5c_1 + 2c_2 + 6)n^2.\]
Setting $c = \max\{1,c_1,c_2\}$, we can write
\[|W_n^{[c_1,c_2]}| \leq 13cn^2.\]
Assume that $n \geq \max\{c,13\}$. Then
\begin{align*}
(\log |W_n|)^{29} & \leq \left( \log 13 + \log c + 2 \log n \right)^{29} \\
				  & \leq (4 \log n)^{29}.
\end{align*}
Thus if we fix $N_0 \geq \max\{c,13\}$ such that $n / (\log^{29}n) > 4^{29}$ for all $n > N_0$, we will have $\tw(W_n^{[c_1,c_2]}) > (\log |W_n^{[c_1,c_2]}|)^{29}$ for all $n > N_0$.  It follows that $\tw(G_n) > (\log |G_n|)^{29}$ for all $n$, as required.
\end{proof}

We now have all the ingredients to prove Theorem \ref{thm:MSO-dichotomy}.

\begin{proof}[Proof of Theorem \ref{thm:MSO-dichotomy}]
The positive direction is immediate from Corollaries \ref{cor:tractable-VC} and \ref{lma:MSO-tractable-deletion}.  For the reverse direction, it suffices to prove hardness of \textsc{MSO-Model-Checking} for graphs; assume that \textsc{SAT} cannot be solved in subexponential time $2^{o(n)}$.  Suppose that $\mathcal{C}_1,\ldots,\mathcal{C}_s$ are classes of graphs such that neither condition (1) nor (2) of Theorem \ref{thm:MSO-dichotomy} holds; it follows that either
\begin{enumerate}
\item at least three of the classes $\mathcal{C}_1,\ldots,\mathcal{C}_s$ do not have bounded vertex cover number, or
\item there exist $i,j \in [s]$ with $i \neq j$ such that $\mathcal{C}_i$ is not at constant vertex cover distance from the class of graphs of maximum degree one and $\mathcal{C}_j$ does not have bounded vertex cover number.
\end{enumerate}
In the first case, assume without loss of generality that $\mathcal{C}_1,\mathcal{C}_2$ and $\mathcal{C}_3$ have unbounded vertex cover number.  By Lemma \ref{lma:3unbddVC} this tells us that $\layer(\mathcal{C}_1,\mathcal{C}_2,\mathcal{C}_3)$ contains arbitrarily large walls; since all the classes are monotone, it follows that all other classes contain arbitrarily large graphs with no edges, so we can conclude that $\layer(\mathcal{C}_1,\ldots,\mathcal{C}_s)$ also contains arbitrarily large walls.  The fact that \textsc{MSO-Model-Checking} is not in XP now follows immediately by Lemma \ref{lma:walls-hard}.

In the second case, Lemma \ref{lma:notconstdel} tells us that, for each $h \in \mathbb{N}$, $\layer(\mathcal{C}_i,\mathcal{C}_j)$ and hence (again using monotonicity) $\layer(\mathcal{C}_1,\ldots,\mathcal{C}_s)$ contains a subdivision of the wall $W_h$ in which each edge is subdivided at most twice.  It therefore follows immediately from Lemma \ref{lma:walls-hard} that \textsc{MSO-Model-Checking} is not in XP.
\end{proof}

\subsection{Implications for graphs encoded with an incidence relation}
\label{sec:graphs}

When a graph $G$ is encoded using an adjacency relation, the Gaifman graph of this structure is isomorphic to $G$ itself, so in this setting we can simply replace ``relational structures'' with ``graphs'' in the statements of Theorems \ref{thm:FO-dichotomy} and \ref{thm:MSO-dichotomy}.  However, this is no longer true if we use an incidence relation to encode the graph.  This distinction is important in the context of our results since the graph properties definable in a given fragment of logic depend on the encoding used: for example, the property of being Hamiltonian is only encodable in MSO when an incidence relation is used.

In this section we give alternative formulations of Theorems \ref{thm:FO-dichotomy} and \ref{thm:MSO-dichotomy} for graphs encoded using an incidence relation, stated in terms of structural properties of the graphs themselves (rather than their encodings).  We observe that, with an incidence relation, the Gaifman graph of a the structure encoding a graph $G = (V,E)$ is isomorphic to 
\[\inc(G) = \left(V \cup E, \{\{ve\}: e \text{ incident with } v\}\right).\]
It therefore suffices to obtain necessary and sufficient conditions on $G$ for $\inc(G)$ to have the properties used in Theorems \ref{thm:FO-dichotomy} and \ref{thm:MSO-dichotomy}.  We begin by giving a necessary condition on $G$ for $\inc(G)$ to be at constant vertex cover distance from a class of graphs of bounded maximum degree.

\begin{lma}\label{lma:inc-deledges}
Let $G = (V,E)$ be a graph, and let $\inc(G)$ be the incidence graph of $G$.  If $W \subseteq V \cup E$ is such that the maximum degree of $\inc(G) \setminus E_{\inc(G)}(W,*)$ is at most $c$, then the number of vertices of degree greater than $c$ in $G$ is at most $2|W|$.
\end{lma}
\begin{proof}
Suppose that $W = V' \cup E'$, where $V' \subseteq V$ and $E' \subseteq E$.  We begin by defining a subset $W'$ of $V$ by setting
\[W' = V' \cup \{\{u,v\}: e = uv \in E'\},\]
and observing that $|W'| \leq |V'| + 2|E'| \leq 2|W|$.  We claim that $\inc(G) \setminus E_{\inc(G)}(W',*)$ also has maximum degree at most $c$.  To see this, note that 
\begin{align*}
E_{\inc(G)} (W',*) & = E_{\inc(G)}(V',*) \cup \bigcup_{e = uv \in E'} E_{\inc(G)}(\{u,v\},*) \\
	& \supseteq E_{\inc(G)}(V',*) \cup \bigcup_{e = uv \in E'} E_{\inc(G)}(e,*) \\
	& = E_{\inc(G)}(V',*) \cup E_{\inc(G)}(E',*) \\
	& = E_{\inc(G)}(W,*).
\end{align*}
We further claim that every vertex with degree in $G$ greater than $c$ must belong to $W'$.  Suppose for a contradiction that there exists $x \in V \setminus W'$ with $d_G(x) > c$.  Since $W'$ contains neither $x$ nor any neighbour of $x$ in $\inc(G)$ (since $W' \subseteq V$), we see that every edge incident with $x$ in $\inc(G)$ survives in $\inc(G) \setminus E_{\inc(G)}(W',*)$.  It follows that the degree of $x$ in $\inc(G) \setminus E_{\inc(G)}(W',*)$ is greater than $c$, giving the required contradiction.

We therefore conclude that every vertex of degree greater than $c$ in $G$ belongs to $W'$, and so the number of such vertices is at most $2|W|$, as required.
\end{proof}

Using this result, we now give a necessary and sufficient condition on $G$ for $\inc(G)$ to have bounded vertex cover number.

\begin{lma}\label{lma:inc-VC}
Let $G = (V,E)$ be a graph.  The incidence graph $\inc(G)$ of $G$ has bounded vertex cover number if and only if the number of non-isolated vertices in $G$ is bounded by a constant.
\end{lma}
\begin{proof}
Suppose first that $U \subseteq V$ is the set of non-isolated vertices in $G$; we will argue that $U$ is in fact a vertex cover in $\inc(G)$.  To see this, consider any edge $f$ in $\inc(G)$; without loss of generality suppose that $f = ve$ with $v \in V$ and $e \in E$.  By definition of $\inc(G)$, $v$ cannot be isolated in $G$, and so $v \in U$.  Since the edge $f$ was chosen arbitrarily, it follows that every edge in $\inc(G)$ has at least one endpoint in $U$, and so $U$ is a vertex cover for $\inc(G)$.  It follows that, if the number of non-isolated vertices in $G$ is bounded by a constant then $\inc(G)$ has bounded vertex cover number.

Conversely, let $W \subseteq V \cup E$ be a vertex cover for $\inc(G)$, so that the maximum degree of $\inc(G) \setminus E_{\inc(G)}(W,*)$ is zero.  It follows from Lemma \ref{lma:inc-deledges} that the number of vertices of degree greater than zero in $G$, that is the number of non-isolated vertices, is at most $2|W|$.  Hence, if the vertex cover number of $\inc(G)$ is bounded by a constant, the number of non-isolated vertices in $G$ must also be bounded by a (different) constant. 
\end{proof}

We use Lemma \ref{lma:inc-deledges} again to characterise those graphs whose incidence graphs are at constance vertex cover distance from the class of graphs of maximum degree one.

\begin{lma}\label{lma:inc-deg1}
Let $G = (V,E)$ be a graph.  The incidence graph $\inc(G)$ of $G$ is at constant vertex cover distance from the class of graphs of maximum degree at most one if and only if $G$ has bounded vertex cover number and a bounded number of vertices of degree at least two.
\end{lma}
\begin{proof}
Let $X \subseteq V$ be a vertex cover for $G$, and let $Y \subseteq V$ be the set of vertices with degree at least two in $G$.  We claim that $H := \inc(G) \setminus E_{\inc(G)}(X \cup Y,*)$ has maximum degree at most one, implying that $\inc(G)$ is at constant vertex cover distance from the class of graphs of maximum degree at most provided that $|X|$ and $|Y|$ are both bounded by constants.  To see that this is true, fix a vertex $z \in \inc(G)$.  If $z \in V$, then either $z \in X \cup Y$, in which case $d_H(z) = 0$, or else $d_H(z) = d_G(z)$ in which case, since $z \notin Y$, we conclude that $d_H(z) \leq 1$.  If $z \in E$ then, since $X$ is a vertex cover for $G$, we know that at least one neighbour of $z$ in $\inc(G)$ belongs to $X$ and so $d_H(z) \leq d_{\inc(G)}(z) - 1 = 1$.

Conversely, suppose that $W \subseteq V \cup E$ is such that $\inc(G) \setminus E_{\inc(G)}(W,*)$ has maximum degree at most one.  By Lemma \ref{lma:inc-deledges}, we know that the number of vertices with degree greater than one is at most $2|W|$ and so, if $\inc(G)$ is at constant vertex cover distance from the class of graphs of maximum degree at most one then $G$ has only a bounded number of vertices of degree at least two.  We further claim that the vertex cover number of $G$ is at most $|W|$.  To see this, note that for every edge $e = uv \in E$, we must either have $e \in W$ or $\{u,v\} \cap W \neq \emptyset$, since otherwise $d_{\inc(G) \setminus E_{\inc(G)}(W,*)}(e) = 2$.  We can therefore obtain a vertex cover $W'$ for $G$ by replacing any $e \in W$ with one of its endpoints (chosen arbitrarily); clearly $|W'| \leq |W|$ and so the vertex cover number of $G$ is at most $W$, as required.  We therefore conclude that, if $\inc(G)$ is at constant vertex cover distance from the class of graphs of maximum degree one, then both the number of vertices with degree at least two in $G$ and the vertex cover number of $G$ must be bounded by constants.
\end{proof}

Finally, we give a characterisation of those graphs whose incidence graphs are at constant vertex cover distance from the class of graphs of maximum degree $c$, for any constant $c$ greater than one.

\begin{lma}\label{lma:inc-constdeg}
Let $G=(V,E)$ be a graph, and let $\mathcal{C}_c$ be the class of graphs of maximum degree at most $c$.  If $c \geq 2$, then the incidence graph $\inc(G)$ of $G$ is at constant vertex cover distance from $\mathcal{C}_c$ if and only if $G$ has a bounded number of vertices of degree greater than $c$.
\end{lma}
\begin{proof}
Let $U$ be the set of vertices in $G$ with degree greater than $c$.  We will first argue that $\inc(G) \setminus E_{\inc(G)}(U,*)$ has maximum degree at most $c$.  Note that $d_{\inc(G)}(e) = 2 \leq c$ for every $e \in E$, and that $d_{\inc(G)}(v) = d_G(v)$ for every $v \in V$.  It follows immediately that deleting from $\inc(H)$ all edges incident with vertices whose degree in $G$ is greater than $c$ results in a graph with maximum degree at most $c$.

Conversely, suppose that $W \subseteq V \cup E$ is such that $\inc(G) \setminus E_{\inc(G)}$ has maximum degree at most $c$.  It follows from Lemma \ref{lma:inc-deledges} that the number of vertices in $G$ with degree greater than $c$ is at most $2|W|$ and hence that if $\inc(G)$ is at constant vertex cover distance from $\mathcal{C}_c$ then $G$ has only a constant number of vertices of degree greater than $c$. 
\end{proof}

Using Lemmas \ref{lma:inc-VC}, \ref{lma:inc-deg1} and \ref{lma:inc-constdeg}, we can now state our dichotomy results for graphs encoded using an incidence relation; we note that the conditions required for tractability are significantly more restrictive in this setting.

\begin{thm}
Let $\mathcal{C}_1,\ldots,\mathcal{C}_s$ be monotone classes of graphs such that, for each $1 \leq i \leq s$, \textsc{FO-Model-Checking} (respectively \textsc{FO-Counting}) is in FPT when restricted to graphs from $\mathcal{C}_i$ encoded using an incidence relation.  Then, assuming FPT $\neq$ W[1], \textsc{FO-Model-Checking} (respectively \textsc{FO-Counting}) is in FPT restricted to graphs from $\layer(\mathcal{C}_1,\ldots,\mathcal{C}_s)$ encoded with an incidence relation if and only if either
\begin{enumerate}
\item there is some $j \in [s]$ such that, for all $i \in [s] \setminus \{j\}$, the number of non-isolated vertices in any element of $\mathcal{C}_i$ bounded by a constant, or
\item there exist constants $c$ and $d$ such that, for each $i \in [s]$, the number of vertices of degree greater than $d$ in any element of $\mathcal{C}_i$ is at most $c$.
\end{enumerate}
\end{thm}

\begin{thm}
Let $\mathcal{C}_1,\ldots,\mathcal{C}_s$ be monotone classes of graphs such that, for each $1 \leq i \leq s$, \textsc{MSO-Model-Checking} (respectively \textsc{MSO-Counting}) is in FPT when restricted to graphs from $\mathcal{C}_i$ encoded using an incidence relation.  \textsc{MSO-Model-Checking} (respectively \textsc{MSO-Counting}) is in FPT restricted to graphs from $\layer(\mathcal{C}_1,\ldots,\mathcal{C}_s)$ encoded with an incidence relation if either
\begin{enumerate}
\item there is some $j \in [s]$ such that, for all $i \in [s] \setminus \{j\}$, the number of non-isolated vertices in any element of $\mathcal{C}_i$ bounded by a constant, or
\item there exist $i,j \in [s]$ such that
\begin{itemize}
\item $\mathcal{C}_i$ and $\mathcal{C}_j$ both have bounded vertex cover number, and
\item the number of vertices of degree at least two in any element of $\mathcal{C}_i \cup \mathcal{C}_j$ is bounded by a constant, and
\item for all $\ell \in [s] \setminus \{i,j\}$, the number of non-isolated vertices in any element of $\mathcal{C}_{\ell}$ is bounded by a constant.
\end{itemize}
\end{enumerate}
Otherwise, unless \textsc{SAT} can be solved in subexponential time $2^{o(n)}$, \textsc{MSO-Model-Checking} is not in XP.
\end{thm}

\section{Conclusions and Future Work}

We have provided a complete characterisation of the settings in which structural properties of individual layers in a multi-layer structure are sufficient to guarantee tractability of \textsc{FO-Model-Checking}, \textsc{FO-Counting}, \textsc{MSO-Model-Checking} and \textsc{MSO-Counting}, provided that the properties of the layers are preserved under deletion of both vertices and edges.  While this monotonicity requirement holds for many structural restrictions of interest (in particular, those restrictions that place some kind of sparsity requirement on the graph), it would be interesting to investigate whether an analogous characterisation holds without this requirement, allowing us to extend the result to include, for example, classes of bounded cliquewidth.

While the dichotomy results here provide a full answer to our original theoretical question, this problem was motivated by the need to exploit structural properties of layers in real-world applications, and unfortunately our results show that, with just a few exceptions, computationally useful structure in each individual layer is not enough to guarantee similarly exploitable structure in the layered system.  It appears that the strength of our hardness results is due in part to the flexibility inherent in our definition of layered graph classes: graphs from each class can be combined using \emph{any} mapping between the vertex sets.  An intriguing direction for further research is therefore whether we can obtain more tractable cases by restricting the way in which layers can be combined; as a very simple example we could impose local conditions on the number of edges from each layer incident at each vertex in the layered graph, as well as enforcing global structural properties in each layer.  A second, related question concerns the likely structural properties of a layered graph when each layer is drawn from a specified class but the mapping between vertices in each layer is determined by a random process: can we characterise situations in which the layered graph has useful structure with high probability, even though this is not guaranteed?

\section*{Acknowledgements}

Jessica Enright is partially supported by EPSRC project EP/P026842/1.  Kitty Meeks is supported by a Royal Society of Edinburgh Personal Research Fellowship, funded by the Scottish Government.  Jessica Ryan is supported by an EPSRC Doctoral Training Account.

\bibliographystyle{amsplain}
\bibliography{ICALP_refs}

\end{document}